\documentclass[aps,showpacs,twocolumn]{revtex4}
\usepackage{epsfig}
\usepackage{amsmath}

\begin{document}
\title{Systematic study of $Z_c^+$ family from quark model's perspective}

\author{Chengrong Deng$^{a,b}{\footnote{crdeng@cqjtu.edu.cn}}$,
        Jialun Ping$^c{\footnote{jlping@njnu.edu.cn, corresponding author}}$,
        Hongxia Huang$^c{\footnote{hxhuang@njnu.edu.cn}}$,
        and Fan Wang$^d{\footnote{fgwang@chenwang.nju.edu.cn}}$}
\affiliation{$^a$School of Science, Chongqing Jiaotong University, Chongqing 400074, P.R. China}
\affiliation{$^b$Department of Physics and Astronomy, University of California, Los Angeles 90095, USA}
\affiliation{$^c$Department of Physics, Nanjing Normal University, Nanjing 210097, P.R. China}
\affiliation{$^d$Department of Physics, Nanjing University, Nanjing 210093, P.R. China}

\begin{abstract}
Inspired by the present experimental status of charged charmonium-like states $Z_c^+$, the tetraquark states $[cu][\bar{c}\bar{d}]$
are systematically studied in a color flux-tube model with a multi-body confinement potential. The investigation indicates that charged
charmonium-like states $Z_c^+(3900)$ or $Z_c^+(3885)$, $Z_c^+(3930)$, $Z_c^+(4020)$ or $Z_c^+(4025)$, $Z_1^+(4050)$, $Z_2^+(4250)$,
and $Z_c^+(4200)$ can be uniformly described as tetraquark states $[cu][\bar{c}\bar{d}]$ with the quantum numbers $n^{2S+1}L_J$ and
$J^P$ of $1^{3}S_1$ and $1^+$, $2^{3}S_1$ and $1^+$, $1^5S_2$ and $2^+$, $1^3P_1$ and $1^-$, $1^5D_1$ and $1^+$, and $1^3D_1$ and $1^+$,
respectively. The predicted lowest charged tetraquark state $[cu][\bar{c}\bar{d}]$ with $0^+$ and $1^1S_0$ has a energy of $3780\pm10$
MeV in the model. The tetraquark states are compact three-dimensional spatial configurations similar to a rugby ball, the higher orbital
angular momentum $L$ between the diquark $[cu]$ and antidiquark $[\bar{c}\bar{d}]$, the more prolate of the states. The multibody color
flux-tube, a collective degree of freedom,  plays an important role in the formation of those charge tetraquark states. However, the two
heavier charged states $Z^+_c(4430)$ and $Z^+_c(4475)$ can not be explained as tetraquark states $[cu][\bar{c}\bar{d}]$ in this model approach.

\end{abstract}


\maketitle

\section{Introduction}

Quantum chromodynamics (QCD) is nowadays widely accepted as the fundamental theory to describe hadrons and their interactions.
Conventional hadrons are composed of either a valence quark $q$ and an antiquark $\bar{q}$ (mesons) or three valence quarks (baryons)
on top of the sea of $q\bar{q}$ pairs and gluons. One of the long standing challenges in hadron physics is to establish  and
classify genuine multiquark states other than conventional hadrons because multiquark states may contain more information about the
low-energy QCD than that of conventional hadrons. In the past several years, a charged charmoniumlike $Z_c^+$ family,
$Z^+_c(4430)$, $Z_1^+(4050)$, $Z_2^+(4250)$, $Z_c^+(3900)$, $Z_c^+(3885)$, $Z_c^+(3930)$, $Z_c^+(4020)$, $Z_c^+(4025)$, $Z_c^+(4475)$
and $Z_c^+(4200)$, has been successively observed by experimental Collaborations~\cite{zc4050,zc3900,zc3885,zc3930,zc4025,zc4020,zc4200,zc4475,zc4430}.
Obviously, those charged charmoniumlike states go beyond conventional $c\bar{c}$-meson picture and prefer to tetraquark systems
$c\bar{c}u\bar{d}$ due to carrying one charge, which provides a good place for testing various phenomenological research methods
of hadron physics. On the theoretical side, a large amount of work has been devoted to describe the internal structure of these charged
states, which has been related to meson-meson molecules~\cite{zc3900molecule,zc4475molecule}, diquark-antidiquark states~\cite{zc3900tetraquark},
hadrocharmonium or Born-Oppenheimer tetraquarks~\cite{hadro-quarkonium}, coupled channel cusps~\cite{cusp}, and kinematic
effects~\cite{kinematiceffects}.

A systematic understanding of the internal structure of these charged states may not only contribute to provide new insights to the
strong interaction dynamics of multiquark systems and low-energy QCD but also provide important information on future experimental
search for the missing higher orbital excitations in the $Z_c^+$ family. This is the goal of the present work. In our approach,
a phenomenological model, color flux-tube model with a multi-body confinement potential instead of a two-body one in traditional
quark model, is employed to explore the properties of excited charged tetraquark states $c\bar{c}u\bar{d}$ systematically. The model
has been successfully applied to the ground states of charged tetraquark states $[Qq][\bar{Q}'\bar{q}']$~($Q, Q'=c,b$ and $q, q'=u,d,s$)
in our previous work~\cite{ground}.

This work is organized as follows: the color flux-tube model and the model parameters are given in Sec. II. The numerical
results and discussions of the charged tetraquark states are presented in Sec. III. A brief summary is given in the last section.

\section{color flux-tube model and parameters}

The details of the color flux-tube model basing on traditional quark models and lattice QCD picture can be found in our previous
work~\cite{flux}, the prominent characteristics of the model are just presented here. The model Hamiltonian for the state
$[cu][\bar{c}\bar{d}]$ is given as follows,
\begin{eqnarray}
H_4 & = & \sum_{i=1}^4 \left(m_i+\frac{\mathbf{p}_i^2}{2m_i}
\right)-T_{C}+\sum_{i>j}^4 V_{ij}+V^C_{min}+V^{C,SL}_{min}, \nonumber\\
V_{ij} & = & V_{ij}^B+V_{ij}^{B,SL}+V_{ij}^{\sigma}+V_{ij}^{\sigma,SL}+V_{ij}^G+V_{ij}^{G,SL}.
\end{eqnarray}
$T_{c}$ is the center-of-mass kinetic energy of the state, $\mathbf{p}_i$ and $m_i$ are the momentum and mass of the $i$-th
quark (antiquark), respectively. The codes of the quarks (antiquarks) $c$ and $u$ ($\bar{c}$ and $\bar{d}$) are assumed to be 1 and
2 (3 and 4), respectively, their positions are denoted as $\mathbf{r}_1$ and $\mathbf{r}_2$ ($\mathbf{r}_3$ and $\mathbf{r}_4$).

The quadratic confinement potential, which is believed to be flavor independent, of the tetraquark state with a diquark-antidiquark
structure has the following form,
\begin{eqnarray}
V^{C}&=&K\left[ (\mathbf{r}_1-\mathbf{y}_{12})^2
+(\mathbf{r}_2-\mathbf{y}_{12})^2+(\mathbf{r}_{3}-\mathbf{y}_{34})^2\right. \nonumber \\
&+&
\left.(\mathbf{r}_4-\mathbf{y}_{34})^2+\kappa_d(\mathbf{y}_{12}-\mathbf{y}_{34})^2\right],
\end{eqnarray}
The positions $\mathbf{y}_{12}$ and $\mathbf{y}_{34}$ are the junctions of two Y-shaped color flux-tube structures. The parameter $K$
is the stiffness of a three-dimension flux-tube, $\kappa_{d}K$ is other compound color flux-tube stiffness.
The relative stiffness parameter $\kappa_{d}$ of the compound flux-tube is~\cite{kappa}
\begin{equation}
\kappa_{d}=\frac{C_{d}}{C_3},
\end{equation}
where $C_{d}$ is the eigenvalue of the Casimir operator associated with the $SU(3)$ color representation $d$ at either end of the color
flux-tube, such as $C_3=\frac{4}{3}$, $C_6=\frac{10}{3}$, and $C_8=3$.

The minimum of the confinement potential $V^C_{min}$ can be obtained by taking the variation of $V^C$ with respect to $\mathbf{y}_{12}$ and
$\mathbf{y}_{34}$,  and it can be expressed as
\begin{eqnarray}
V^C_{min}&=& K\left(\mathbf{R}_1^2+\mathbf{R}_2^2+
\frac{\kappa_{d}}{1+\kappa_{d}}\mathbf{R}_3^2\right),
\end{eqnarray}
The canonical coordinates $\mathbf{R}_i$ have the following forms,
\begin{eqnarray}
\mathbf{R}_{1} & = &
\frac{1}{\sqrt{2}}(\mathbf{r}_1-\mathbf{r}_2),~
\mathbf{R}_{2} =  \frac{1}{\sqrt{2}}(\mathbf{r}_3-\mathbf{r}_4), \nonumber \\
\mathbf{R}_{3} & = &\frac{1}{ \sqrt{4}}(\mathbf{r}_1+\mathbf{r}_2
-\mathbf{r}_3-\mathbf{r}_4), \\
\mathbf{R}_{4} & = &\frac{1}{ \sqrt{4}}(\mathbf{r}_1+\mathbf{r}_2
+\mathbf{r}_3+\mathbf{r}_4). \nonumber
\end{eqnarray}
The use of $V^C_{min}$ can be understood here that the gluon field readjusts immediately to its minimal configuration.
It is worth emphasizing that the confinement $V^C_{min}$ is a multi-body interaction in a multiquark state rather than
the sum of many pairwise confinement interactions,
\begin{eqnarray}
V^C=\sum_{i<j}\lambda_i\cdot\lambda_j r^n_{ij},
\end{eqnarray}
in Isgur-Karl quark model and chiral quark model with $n=1$ or 2.

The central parts of one-boson-exchange $V_{ij}^{B}$ and $\sigma$-meson exchange $V_{ij}^{\sigma}$ only occur between
$u$ and $\bar{d}$, and that of one-gluon-exchange $V_{ij}^G$ is universal. $V_{ij}^{B}$, $V_{ij}^{\sigma}$ and $V_{ij}^{G}$
take their standard forms and are listed in the following,
\begin{eqnarray}
V_{ij}^{B} & = & V^{\pi}_{ij} \sum_{k=1}^3 \mathbf{F}_i^k
\mathbf{F}_j^k+V^{K}_{ij} \sum_{k=4}^7\mathbf{F}_i^k\mathbf{F}_j^k \nonumber\\
&+&V^{\eta}_{ij} (\mathbf{F}^8_i \mathbf{F}^8_j\cos \theta_P
-\sin \theta_P),\\
V^{\chi}_{ij} & = &
\frac{g^2_{ch}}{4\pi}\frac{m^3_{\chi}}{12m_im_j}
\frac{\Lambda^{2}_{\chi}}{\Lambda^{2}_{\chi}-m_{\chi}^2}
\mathbf{\sigma}_{i}\cdot
\mathbf{\sigma}_{j} \nonumber \\
& \times &\left( Y(m_\chi r_{ij})-
\frac{\Lambda^{3}_{\chi}}{m_{\chi}^3}Y(\Lambda_{\chi} r_{ij})
\right),  \\
V_{ij}^{G} & = & {\frac{\alpha_{s}}{4}}\mathbf{\lambda}^c_{i}
\cdot\mathbf{\lambda}_{j}^c\left({\frac{1}{r_{ij}}}-
{\frac{2\pi\delta(\mathbf{r}_{ij})\mathbf{\sigma}_{i}\cdot
\mathbf{\sigma}_{j}}{3m_im_j}}\right), \\
V^{\sigma}_{ij} & = &-\frac{g^2_{ch}}{4\pi}
\frac{\Lambda^{2}_{\sigma}m_{\sigma}}{\Lambda^{2}_{\sigma}-m_{\sigma}^2}
\left( Y(m_\sigma r_{ij})-
\frac{\Lambda_{\sigma}}{m_{\sigma}}Y(\Lambda_{\sigma}r_{ij})
\right). \nonumber  \\
\end{eqnarray}
Where $\chi$ stands for $\pi$, $K$ and $\eta$, $Y(x)=e^{-x}/x$. The symbols $\mathbf{F}$, $\boldsymbol{\lambda}$ and
$\boldsymbol{\sigma}$ are the flavor $SU(3)$, color SU(3) Gell-Mann and spin $SU(2)$ Pauli matrices, respectively.
$\theta_P$ is the mixing angle between $\eta_1$ and $\eta_8$ to give the physical $\eta$ meson.  $g^2_{ch}/4\pi$ is
the chiral coupling constant. $\alpha_s$ is the running strong coupling constant and takes the following form~\cite{vijande},
\begin{equation}
\alpha_s(\mu_{ij})=\frac{\alpha_0}{\ln\left((\mu_{ij}^{2}+\mu_0^2)/\Lambda_0^2\right)},
\end{equation}
where $\mu_{ij}$ is the reduced mass of two interacting particles $q_i$ (or $\bar{q}_i$) and $q_j$ (or $\bar{q}_j$). $\Lambda_0$,
$\alpha_0$ and $\mu_0$ are model parameters. The function $\delta(\mathbf{r}_{ij})$ in $V_{ij}^G$ should be regularized~\cite{weistein},
\begin{equation}
\delta(\mathbf{r}_{ij})=\frac{1}{4\pi r_{ij}r_0^2(\mu_{ij})}e^{-r_{ij}/r_0(\mu_{ij})},
\end{equation}
where $r_0(\mu_{ij})=\hat{r}_0/\mu_{ij}$, $\hat{r}_0$ is a model parameter.

The diquark $[cu]$ and antidiquark $[\bar{c}\bar{d}]$ can be considered as compound objects $\bar{Q}$
and $Q$ with no internal orbital excitation, and the angular excitation $L$ are assumed to occur
only between $Q$ and $\bar{Q}$ in the present work and the parity of the state $[cu][\bar{c}\bar{d}]$
is therefore simply related to $L$ as $P=(-1)^L$. In this way, the state $[cu][\bar{c}\bar{d}]$
has lower energy than that of the states with additional internal orbital excitation in $Q$ and $\bar{Q}$.
In order to facilitate numerical calculations, the spin-orbit interactions are approximately assumed
to just take place between compound objects $\bar{Q}$ and $Q$, which is consistence with the work~\cite{ls}.
The related interactions can be presented as follows
\begin{eqnarray}
V_{12,34}^{G,LS} & \approx& {\frac{\alpha_{s}}{4}}\mathbf{\lambda}^{\bar{c}}_{12}
\cdot\mathbf{\lambda}^c_{34}{\frac{1}{8M_{12}M_{34}}}\frac{3}{X^3}
\mathbf{L}\cdot\mathbf{S},\\
V^{\sigma,LS}_{12,34}&\approx&-\frac{g_{ch}^2}{4\pi}
\frac{\Lambda^{2}_{\sigma}}{\Lambda^{2}_{\sigma}-m_{\sigma}^2}
\frac{m_{\sigma}^3}{2M_{12}M_{34}}\mathbf{L}\cdot\mathbf{S}\nonumber\\
& \times &\left( G(m_\sigma X)-
\frac{\Lambda^3_{\sigma}}{m^3_{\sigma}}G(\Lambda_{\sigma}X)
\right),\\
V_{12,34}^{C,LS} &\approx& \frac{K}{8M_{12}M_{34}}\frac{\kappa_d}{1+\kappa_d}
\mathbf{L}\cdot\mathbf{S}.
\end{eqnarray}
where $M_{12}=M_{34}=m_c+m_{u,d}$, $G(x)=Y(x)(\frac{1}{x}+\frac{1}{x^2})$, and $S$ stands for the total spin angular momentum
of the tetraquark state $[cu][\bar{c}\bar{d}]$.

The model parameters are determined as follows. The mass parameters $m_{\pi}$, $m_K$ and $m_{\eta}$ in the interaction
$V^B_{ij}$ are taken their experimental values, namely, $m_{\pi}=0.7$ fm$^{-1}$, $m_K=2.51$ fm$^{-1}$ and $m_{\eta}=2.77$
fm$^{-1}$. The cutoff parameters take the values, $\Lambda_{\pi}=\Lambda_{\sigma}=4.20$ fm$^{-1}$ and $\Lambda_{\eta}=\Lambda_{K}=5.20$
fm$^{-1}$, the mixing angle $\theta_{P}=-15^{o}$~\cite{vijande}. The mass parameter $m_{\sigma}$ in the interaction
$V_{ij}^{\sigma}$ is determined through the PCAC relation $m^2_{\sigma}\approx m^2_{\pi}+4m^2_{u,d}$~\cite{masssigma},
$m_{u,d}=280$ MeV and $m_{\sigma}=2.92$ fm$^{-1}$.  The chiral coupling constant $g_{ch}$ is determined from the $\pi NN$
coupling constant through
\begin{equation}
\frac{g_{ch}^2}{4\pi}=\left(\frac{3}{5}\right)^2\frac{g_{\pi NN}^2}{4\pi}
\frac{m_{u,d}^2}{m_N^2}=0.43.
\end{equation}
The other adjustable parameters and their errors are determined by fitting the masses of the ground states of mesons
using Minuit program, which are shown in Table I. The mass spectrum of the ground states of mesons, which is listed in Tale II,
can be obtained by solving the two-body Schr\"{o}dinger equation
\begin{eqnarray}
(H_2-E_{2})\Phi_{IJ}^{Meson}=0.
\end{eqnarray}

The mass error of mesons $\Delta E_2$ introduced by the parameter uncertainty $\Delta x_i$ can be calculated by the formula of
error propagation,
\begin{eqnarray}
\Delta H_2 &= & \sum_{i=1}^{8}\left|{\frac{\partial{H_2}}{\partial{x_i}}}\right|\Delta x_i, \\
\Delta E_2 & \approx & \left <\Phi_{IJ}^{Meson}\left|\Delta
H_2\right|\Phi_{IJ}^{Meson}\right >.
\end{eqnarray}
where $x_i$ and $\Delta x_i$ represent the $i$-th adjustable parameter and it's error, respectively, which are listed
in Table I.

\begin{table}[ht]
\caption{Adjustable model parameters.
(units: $m_s$, $m_c$, $m_b$, $\mu_0$, $\Lambda_0$, MeV; $K$, MeV$\cdot$fm$^{-2}$;
$r_0$, MeV$\cdot$fm; $\alpha_0$, dimensionless)\label{para}}
\begin{tabular}{cccccc}

\hline

Parameters     & ~~~~~$x_i$~~~~~    &  ~~$\Delta x_i$~~  &    Parameters   & ~~~~~$x_i$~~~~~   &  ~$\Delta x_i$~ \\
$m_{s}$        &      511.78        &      0.228         &    $\alpha_0$   &      4.554        &      0.018      \\
$m_{c}$        &      1601.7        &      0.441         &    $\Lambda_0$  &      9.173        &      0.175      \\
$m_{b}$        &      4936.2        &      0.451         &    $\mu_0$      &      0.0004       &      0.540      \\
$K$            &      217.50        &      0.230         &    $r_0$        &      35.06        &      0.156      \\

\hline

\end{tabular}
%\end{table}
%\begin{table}
\caption{Ground state meson spectra, unit in MeV.\label{meson}}
\begin{tabular}{ccccccccccccc}

\hline

States         &  ~~~$E_2$~~~ & $\Delta E_2$ &   ~~PDG~~    &     States         &  ~~~$E_2$~~~ & $\Delta E_2$   &    ~~PDG~~  \\
$\pi$          &     142      &       26     &     139      &     $\eta_c$       &     2912     &       5        &     2980    \\
$K$            &     492      &       20     &     496      &     $J/\Psi$       &     3102     &       4        &     3097    \\
$\rho$         &     826      &       4      &     775      &     $B^0$          &     5259     &       5        &     5280    \\
$\omega$       &     780      &       4      &     783      &     $B^*$          &     5301     &       4        &     5325    \\
$K^*$          &     974      &       4      &     892      &     $B_s^0$        &     5377     &       5        &     5366    \\
$\phi$         &     1112     &       4      &     1020     &     $B_s^*$        &     5430     &       4        &     5416    \\
$D^{\pm}$      &     1867     &       8      &     1880     &     $B_c$          &     6261     &       7        &     6277    \\
$D^*$          &     2002     &       4      &     2007     &     $B_c^*$        &     6357     &       4        &     ...     \\
$D_s^{\pm}$    &     1972     &       9      &     1968     &     $\eta_b$       &     9441     &       8        &     9391    \\
$D_s^*$        &     2140     &       4      &     2112     &     $\Upsilon(1S)$ &     9546     &       5        &     9460    \\

\hline

\end{tabular}
\end{table}

\section{numerical results and discussions}

Within the framework of the diquark-antidiquark configuration, the wave function of the state $[cu][\bar{c}\bar{d}]$
can be written as a sum of the following direct products of color $\chi_c$, isospin $\eta_I$, spin $\eta_s$ and spatial
$\phi$ terms,
\begin{eqnarray}
\Phi^{[cu][\bar{c}\bar{d}]}_{IM_IJM_J} &=&
\sum_{\alpha}\xi_{\alpha}\left[ \left[
\left[\phi_{l_am_a}^G(\mathbf{r})\chi_{s_a}\right]^{[cu]}_{j_a}
\left[\phi_{l_bm_b}^G(\mathbf{R})\right.\right.\right.\nonumber\\
& \times & \left.\left.\left.\chi_{s_b}\right]^{[\bar{c}\bar{d}]}_{j_b}
\right ]_{J_{ab}}^{[cu][\bar{c}\bar{d}]}
F_{LM}(\mathbf{X})\right]^{[cu][\bar{c}\bar{d}]}_{JM_J}\\
& \times &
\left[\eta_{I_a}^{[cu]}\eta_{I_b}^{[\bar{c}\bar{d}]}\right]_{IM_I}^{[cu][\bar{c}\bar{d}]}
\left[\chi_{c_a}^{[cu]}\chi_{c_b}^{[\bar{c}\bar{d}]}\right]_{CW_C}^{[cu][\bar{c}\bar{d}]},
\nonumber
\end{eqnarray}
In which $\mathbf{r}$, $\mathbf{R}$ and $\mathbf{X}$ are relative spatial coordinates,
\begin{eqnarray}
\mathbf{r}&=&\mathbf{r}_1-\mathbf{r}_2,~~~\mathbf{R}=\mathbf{r}_3-\mathbf{r}_4 \nonumber\\
\mathbf{X}&=&\frac{m_1\mathbf{r}_1+m_2\mathbf{r}_2}{m_1+m_2}-\frac{m_3\mathbf{r}_3+m_4\mathbf{r}_4}{m_3+m_4}.
\end{eqnarray}
The other details of the construction of the wave function can be found in our previous work~\cite{ground}.
Subsequently, the converged numerical results can be obtained by solving the four-body Schr\"{o}dinger equation
\begin{eqnarray}
(H_4-E_4)\Phi^{[cu][\bar{c}\bar{d}]}_{IM_IJM_J}=0.
\end{eqnarray}
with the Rayleigh-Ritz variational principle.

\begin{table*}
\caption{The energy $E_4+\Delta E_4$ and rms $\langle\mathbf{r}^2\rangle^{\frac{1}{2}}$, $\langle\mathbf{R}^2\rangle^{\frac{1}{2}}$ and
$\langle\mathbf{X}^2\rangle^{\frac{1}{2}}$ of charged tetraquark states $[cu][\bar{c}\bar{d}]$ with $J^{P}$ and $n^{2S+1}L_J$, unit of
energy: MeV and unit of rms: fm.}
\begin{tabular}{ccccccccccccccccccccc}

\hline

$J^{P}$             &$0^{+}$      &$0^{-}$      &$0^{+}$      &$1^{+}$       &$1^{+}$       &$1^{-}$      &$1^{-}$      &$1^{-}$      &$1^{+}$      &$1^{+}$      \\
$n^{2S+1}L_J$       &$1^{1}S_0$   &$1^{3}P_0$   &$1^{5}D_0$   &$1^{3}S_1$    &$2^{3}S_1$    &$1^{1}P_1$   &$1^{3}P_1$   &$1^{5}P_1$   &$1^{3}D_1$   &$1^{5}D_1$   \\
~$E_4\pm\Delta E_4$~&$3782\pm12$~ &~$4097\pm8$~ &~$4274\pm7$~ &~$3858\pm10$~ &~$3950\pm10$~ &~$4075\pm8$~ &~$4097\pm8$~ &~$4153\pm7$~ &~$4235\pm7$~ &~$4273\pm7$~ \\
$\langle\mathbf{r}^2\rangle^{\frac{1}{2}}$ &0.85&0.96&1.01&0.90&0.92&0.94&0.96&1.00&0.98&1.01\\
$\langle\mathbf{R}^2\rangle^{\frac{1}{2}}$ &0.85&0.96&1.01&0.90&0.92&0.94&0.96&1.00&0.98&1.01\\
$\langle\mathbf{X}^2\rangle^{\frac{1}{2}}$ &0.42&0.85&1.12&0.48&0.66&0.85&0.85&0.92&1.10&1.12\\

$J^{P}$             &$1^{-}$      &$2^{+}$      &$2^{-}$      &$2^{-}$       &$2^{+}$       &$2^{+}$      &$2^{+}$      &$2^{-}$      &$2^{-}$      &$3^{-}$      \\
$n^{2S+1}L_J$       &$1^{5}F_1$   &$1^{5}S_2$   &$1^{3}P_2$   &$1^{5}P_2$    &$1^{1}D_2$    &$1^{3}D_2$   &$1^{5}D_2$   &$1^{3}F_2$   &$1^{5}F_2$   &$1^{5}P_3$   \\
$E_4\pm\Delta E_4$  &$4387\pm7$   &$4001\pm7$   &$4096\pm8$   &$4152\pm7$    &$4212\pm8$    &$4235\pm7$   &$4273\pm7$   &$4354\pm7$   &$4387\pm7$   &$4150\pm7$   \\
$\langle\mathbf{r}^2\rangle^{\frac{1}{2}}$ &1.02&1.03&0.96&1.00&0.95&0.98&1.01&0.99&1.02&1.00\\
$\langle\mathbf{R}^2\rangle^{\frac{1}{2}}$ &1.02&1.03&0.96&1.00&0.95&0.98&1.01&0.99&1.02&1.00\\
$\langle\mathbf{X}^2\rangle^{\frac{1}{2}}$ &1.30&0.57&0.85&0.92&1.09&1.10&1.12&1.30&1.30&0.92\\

$J^{P}$             &$3^{+}$      &$3^{+}$      &$3^{-}$      &$3^{-}$       &$3^{-}$       &$4^{+}$      &$4^{-}$      &$4^{-}$      &$5^{-}$     \\
$n^{2S+1}L_J$       &$1^{3}D_3$   &$1^{5}D_3$   &$1^{1}F_3$   &$1^{3}F_3$    &$1^{5}F_3$    &$1^{5}D_4$   &$1^{3}F_4$   &$1^{5}F_4$   &$1^{5}F_5$  \\
$E_4\pm\Delta E_4$  &$4234\pm7$   &$4272\pm7$   &$4332\pm8$   &$4353\pm7$    &$4386\pm7$    &$4274\pm7$   &$4353\pm7$   &$4387\pm7$   &$4387\pm7$  \\
$\langle\mathbf{r}^2\rangle^{\frac{1}{2}}$ &0.98&1.01&0.96&0.99&1.02&1.01&0.99&1.02&1.02\\
$\langle\mathbf{R}^2\rangle^{\frac{1}{2}}$ &0.98&1.01&0.96&0.99&1.02&1.01&0.99&1.02&1.02\\
$\langle\mathbf{X}^2\rangle^{\frac{1}{2}}$ &1.10&1.12&1.27&1.30&1.30&1.12&1.30&1.30&1.30\\

\hline

\end{tabular}
\end{table*}

The energies $E_4\pm\Delta E_4$ of the charged states $[cu][\bar{c}\bar{d}]$ with $n^{2S+1}L_J$ and $J^{P}$ under the assumptions of
$S=0,...,2$ and $L=0,...,3$ are systematically calculated and presented in Table III. The mass error of the states $\Delta E_4$ can be
calculated just as $\Delta E_2$, they are around several MeV except for that of the state $1^1S_0$. The spin-orbit interactions are
extremely weak, less than 2 MeV, therefore the energies of excited states with the same $L$ and $S$ but different $J$ are almost
degenerate, see the energies of the excited states with $1^5D_0$, $1^5D_1$ $1^5D_2$, $1^5D_3$ and $1^5D_4$ in Table III, which is
consistent with the conclusion of the work~\cite{spin-orbit}. Other spin-related interactions are stronger and bring about a larger
energy difference than spin-orbital interactions, especially for the ground states with $1^1S_0$, $1^3S_1$ and $1^5S_2$. The energy
difference among excited states mainly comes from the kinetic energy and confinement potential, which are proportional to the relative
orbital excitation $L$. However, the relative kinetic energy between two clusters $[cu]$ and $[\bar{c}\bar{d}]$ is inversely proportional
to $\langle\mathbf{X}^2\rangle$ while confinement potential is proportional to $\langle\mathbf{X}^2\rangle$ so that they compete each
other to reach an optimum balance.

The rms $\langle\mathbf{r}^2\rangle^{\frac{1}{2}}$, $\langle\mathbf{R}^2\rangle^{\frac{1}{2}}$ and $\langle\mathbf{X}^2\rangle^{\frac{1}{2}}$
stand for the size of the diquark $[cu]$, the antidiquark $[\bar{c}\bar{d}]$ and the distance between the two clusters, respectively, which
are also calculated and listed in Table III. One can find that the diquark $[cu]$ and antidiquark $[\bar{c}\bar{d}]$ share the same size
in every $Z^+_c$ state. The sizes of the diquark $[cu]$ and antidiquark $[\bar{c}\bar{d}]$ ($\langle\mathbf{r}^2\rangle^{\frac{1}{2}}$ and $\langle\mathbf{R}^2\rangle^{\frac{1}{2}}$) are mainly determined by the total spin $S$, the relative orbital excitation $L$ of the states
has a minor effect on them. However, the sizes do not vary largely with the total spin $S$, especially for higher orbital excited states.
So the diquark $[cu]$ and antidiquark $[\bar{c}\bar{d}]$ are rather rigid against the rotation. For examples, the sizes of the two groups $1^1S_0$-$1^3S_1$-$1^5S_2$ and $1^1F_3$-$1^3F_3$-$1^5F_3$ changes gradually with the total spin $S$, 0.85-0.90-1.03 fm and 0.96-0.99-1.02 fm,
respectively. And the sizes of the two groups $1^1S_0$-$1^1P_1$-$1^1D_2$-$1^1F_3$ and $1^3S_1$-$1^3P_1$-$1^3D_1$-$1^3F_2$ vary slightly with
relative orbital excitation $L$, 0.85-0.94-0.95-0.96 fm and 0.90-0.96-0.98-0.99 fm, respectively. On the contrary, the distance between the
diquark $[cu]$ and antidiquark $[\bar{c}\bar{d}]$ ($\langle\mathbf{X}^2\rangle^{\frac{1}{2}}$) changes remarkably with the relative orbital
excitation $L$ between the two clusters and is irrelevant to the total spin of the system, see the sizes of $1^3S_1$-$1^3P_1$-$1^3D_1$-$1^3F_2$
and $1^1S_0$-$1^3S_1$-$1^5S_2$ in Table III. The sizes of the diquark $[cu]$, antidiquark $[\bar{c}\bar{d}]$ and the distance between the
two clusters are helpful to understand the changing tendency of energies of charged states $Z_c^+$ with quantum numbers $S$ and $L$.

\begin{table}
\caption{The average distances $\langle\mathbf{r}_{ij}^2\rangle^{\frac{1}{2}}$ of the states $[cu][\bar{c}\bar{d}]$ with $1^1S_0$, $1^1P_1$,
$1^1D_2$, and $1^1F_3$, $\mathbf{r}_{ij}=\mathbf{r}_i-\mathbf{r}_j$, unit in fm.}
\begin{tabular}{cccccccccc} \hline
$n^{2S+1}L_J$&$\langle\mathbf{r}_{12}^2\rangle^{\frac{1}{2}}$&$\langle\mathbf{r}_{34}^2\rangle^{\frac{1}{2}}$&$\langle\mathbf{r}_{24}^2\rangle^{\frac{1}{2}}$
           &$\langle\mathbf{r}_{13}^2\rangle^{\frac{1}{2}}$&$\langle\mathbf{r}_{14}^2\rangle^{\frac{1}{2}}$&$\langle\mathbf{r}_{23}^2\rangle^{\frac{1}{2}}$
           &$\langle\mathbf{X}^2\rangle^{\frac{1}{2}}$\\
$1^1S_0$   &   0.85         &    0.85        &    1.11        &    0.46          &     0.85      &    0.85     &    0.42     \\
$1^1P_1$   &   0.94         &    0.94        &    1.41        &    0.87          &     1.17      &    1.17     &    0.85     \\
$1^1D_2$   &   0.95         &    0.95        &    1.59        &    1.11          &     1.37      &    1.37     &    1.09     \\
$1^1F_3$   &   0.96         &    0.96        &    1.72        &    1.28          &     1.52      &    1.52     &    1.27     \\
\hline
\end{tabular}
\end{table}

In order to make clear the spatial configuration of charged states $[cu][\bar{c}\bar{d}]$, the distances in four states between any
two particles are given in Table IV. The ground state ($1^1S_0$ and $1^+$) of charged tetraquark $[cu][\bar{c}\bar{d}]$ possesses a
three-dimensional spatial configuration due to the competition of the confinement and the kinetic energy of the systems~\cite{ground},
which is similar to a rugby ball. The diquark $[cu]$ and antidiqurk $[\bar{c}\bar{d}]$ in the ground state have a large overlap because
of the small $\langle\mathbf{X}^2\rangle^{\frac{1}{2}}$, so the picture of the diqurk or antiquark is not extremely distinct. However,
all distances except for the sizes of the diquark and antidiquark ($\langle\mathbf{r}_{12}^2\rangle^{\frac{1}{2}}$ and 
$\langle\mathbf{r}_{34}^2\rangle^{\frac{1}{2}}$) evidently augment with the increasing of the orbital angular momentum $L$ in the excited
states, see Table IV, which means that the picture of the diquark or antidiquark is more and more clear with the raising of the
orbital angular momentum $L$. The spatial configuration of the excited states is still similar to a rugby ball, the higher orbital angular
momentum $L$, the more prolate of the shape of the excited states. The multibody color flux-tube basing on lattice QCD picture, a collective
degree of freedom, plays an important role in the formation of these charged tetraquark states, it should therefore be the dynamical
mechanism of the tetraquark systems.

\begin{table}[ht]
\caption{$Z^+_c$ states observed in experiments and their possible candidates in the color flux-tube model.}
\begin{tabular}{ccccccccc}

\hline

                            &      ~Experiment~              &         &                &    Model  &                \\
State                       &        Mass, MeV               &~$J^{P}$~&    Mass, MeV   &  $J^{P}$  &  $n^{2S+1}L_J$ \\
$Z_1^+(4050)$~\cite{zc4050} & $4051^{+14+20}_{-14-41}$       & $?^{?}$ &   $4075\pm8$   &  $1^{-}$  &   $1^1P_1$     \\
$Z_2^+(4250)$~\cite{zc4050} & $4248^{+44+180}_{-29-35}$      & $?^{?}$ &   $4273\pm7$   &  $1^{+}$  &   $1^5D_1$     \\
$Z_c^+(3900)$~\cite{zc3900} & $3899.0^{+3.6+4.9}_{-3.6-4.9}$ & $?^{?}$ &   $3858\pm10$  &  $1^{+}$  &   $1^3S_1$     \\
$Z_c^+(3885)$~\cite{zc3885} & $3883.9^{+1.5+4.2}_{-1.5-4.2}$ & $1^{+}$ &   $3858\pm10$  &  $1^{+}$  &   $1^3S_1$     \\
$Z_c^+(3930)$~\cite{zc3930} & $3929^{+5+2}_{-5-2}$           & $1^{+}$ &   $3950\pm10$  &  $1^{+}$  &   $2^3S_1$     \\
$Z_c^+(4025)$~\cite{zc4025} & $4026.3^{+2.6+3.7}_{-2.6-3.7}$ & $?^{?}$ &   $4001\pm7$   &  $2^{+}$  &   $1^5S_2$     \\
$Z_c^+(4020)$~\cite{zc4020} & $4022.9^{+0.8+2.7}_{-0.8-2.7}$ & $?^{?}$ &   $4001\pm7$   &  $2^{+}$  &   $1^5S_2$     \\
$Z_c^+(4200)$~\cite{zc4200} & $4196^{+36+17}_{-29-13}$       & $?^{?}$ &   $4235\pm7$   &  $1^{+}$  &   $1^3D_1$     \\
$Z_c^+(4475)$~\cite{zc4475} & $4475^{+22+28}_{-22-11}$       & $1^{+}$ &      ...       &    ...    &     ...        \\
$Z_c^+(4430)$~\cite{zc4430} & $4433^{+2+4}_{-2-4}$           & $1^{+}$ &      ...       &    ...    &     ...        \\
\hline

\end{tabular}
\end{table}

Next, let us turn to argue the properties of the charged states $Z^+_c$ observed in experiments and their possible candidates
in the color flux-tube model, which are presented in Table IV. It can be seen from the Table IV that the spin and parity of the
$Z^+_c(3900)$ are still unclear up to now. The $Z^+_c(3900)$ may correspond to the same state as the $Z^+_c(3885)$ with $1^+$
~\cite{zc3885}. The charged state $[cu][\bar{c}\bar{d}]$ with $1^+$ and $1^3S_1$ has a mass of $3858\pm10$ MeV in the color
flux-tube model, which is very close to those of the two charged states $Z^+_c(3885)$ and $Z^+_c(3900)$. It can not be excluded
that the main component of $Z^+_c(3885)$ and $Z^+_c(3900)$ is the state $[cu][\bar{c}\bar{d}]$ with $1^+$ and $1^3S_1$, which
is supported by many theoretical work~\cite{zc3900tetraquark}. The radial excited state $2^3S_1$ has a mass of $3950\pm10$ MeV,
which is extremely close to that of $Z_c^+(3930)$. It is possible to identify $Z_c^+(3930)$ as the tetraquark state $[cu][\bar{c}\bar{d}]$
with $1^+$ and $2^3S_1$. In other words, the $Z^+_c(3930)$ is the first radial excited state of the $Z^+_c(3900)$ in the color
flux-tube model.  The pair $Z^+_c(4020)$ and $Z^+_c(4025)$ show up with a similar mass (slightly above $D^*\bar{D}^*$ threshold).
They might therefore be the same resonance, their spin and parity are unclear. QCD sum rule identified the $Z^+_c(4020)$ and
$Z^+_c(4025)$ as a tetraquark state $[cu][\bar{c}\bar{d}]$ with  $1^+$~\cite{zgwang}, the same approach also favored a tetraquark
state but with different quantum numbers $2^+$ and $^5S_2$~\cite{sumrule2+}. In our calculations, the nearest tetraquark state
$[cu][\bar{c}\bar{d}]$ to the $Z^+_c(4020)$ or $Z^+_c(4025)$ occupies quantum numbers $2^+$ and $1^5S_2$. The tetraquark states
$[cu][\bar{c}\bar{d}]$ with $1^-$ and $1^1P_1$ and $1^+$ and $1^5D_1$ have the energies of $4075\pm8$ MeV and $4273\pm7$ MeV,
respectively, which are consistent with those of $Z^+_1(4050)$ and $Z^+_2(4250)$. So the two states may be assigned as the tetraquark
states $[cu][\bar{c}\bar{d}]$ with $1^-$ and $1^1P_1$ and $1^+$ and$1^5D_1$, respectively, in the color flux-tube model. The newly
observed $Z^+_c(4200)$ prefers $1^+$, which can be described as the tetraquark state $[cu][\bar{c}\bar{d}]$ with $1^+$ and $1^3D_1$
in the color flux-tube model. The study of the three-point function sum rules on this state gives a support to the tetraquark
interpretation~\cite{wchen4200}. Of cause, it seems difficult to rule out other two possibilities of $2^+$ and $1^1D_2$ and $2^+$
and $1^3D_2$ in the model. The $Z^+_c(4430)$ is the first charged state, the $J^P$ of the state is determined unambiguously to be
$1^+$, the $Z^+_c(4475)$ favors the spin-parity $1^+$ over other hypotheses~\cite{zc4475}. Due to the heavy of the diquark and
antidiquark, the energy of radial excitation between the diquark and antidiquark is too small to make the tetraquark state
$[cu][\bar{c}\bar{d}]$ reach the energy above 4400 MeV. The internal excited states of the diquark and/or antidiquark are
needed to account for the heavy charged states, which is left for future. The alternative configuration for the two states may be
meson-meson molecular states, which is suggested by several theoretical methods~\cite{zc4475molecule}.

From the above analysis and Table III, we can see that the most of the low energy theoretical states can be matched with the
experimental ones. One of the exception is the state with $0^+$ and $1^1S_0$, which has a mass of $3780\pm10$ MeV.
The experimental search of the $\eta_c$-like charged state will give a crucial test of the present approach. Our calculation also
suggests that there are two negative parity states around 4100 MeV. More information on the the states around this energy is expected.

The model assignments of the $Z^+_c$ states are completed just in term of the proximity to the experimental masses, the more
stringent check of the assignment is to study the decay properties of the states. These charged states should eventually decay into
several color singlet mesons due to their high energy. In the course of the decay, the color flux-tube structure should break down
first which leads to the collapses of the three-dimension spatial configuration, and then through the recombination of the color
flux tubes the particles of decay products formed. The decay widths of the charged states $[cu][\bar{c}\bar{d}]$ are determined
by the transition probability of the breakdown and recombination of color flux tubes. The calculations are in progress. This
decay mechanism is similar to compound nucleus decay and therefore should induce a resonance, which we called it as ``color confined,
multiquark resonance" state before~\cite{resonance}.

\section{summary}

The charged tetraquark states $[cu][\bar{c}\bar{d}]$ are systematically  investigated from the perspective of the color flux-tube model
with a four-body confinement potential. The investigation demonstrates that the charged charmoniumlike states $Z_c^+(3900)$ or $Z_c^+(3885)$,
$Z_c^+(3930)$, $Z_c^+(4020)$ or $Z_c^+(4025)$, $Z_1^+(4050)$, $Z_2^+(4250)$, and $Z_c^+(4200)$ can be uniformly identified as tetraquark
states $[cu][\bar{c}\bar{d}]$ with the quantum numbers $1^{3}S_1$ and $1^+$, $2^{3}S_1$ and $1^+$, $1^5S_2$ and $2^+$, $1^3P_1$ and $1^-$,
$1^5D_1$ and $1^+$, and $1^3D_1$ and $1^+$, respectively, in the color flux-tube model. The predicted lowest charged tetraquark state
$[cu][\bar{c}\bar{d}]$ with $0^+$ and $1^1S_0$ has a mass of $3780\pm10$ MeV in the color flux-tube model. The model predictions would
shed light on looking for possible charmoniumlike charged states in the future at the BESIII, LHCb and Belle-II. They favor three-dimensional
spatial structures which is similar to a rugby ball,the higher orbital angular momentum $L$, the more prolate of the shape of the states.
Those charged charmoniumlike states may be so-called ``color confined, multiquark resonance". However, the two heavier charged states
$Z^+_c(4430)$ and $Z^+_c(4475)$ can not be described as tetraquark states $[cu][\bar{c}\bar{d}]$ in the color flux-tube model.

The multibody color flux-tube is a collective degree of freedom, which acts as a dynamical mechanism and plays an important role in the
formation and decay of those compact states. Just as colorful organic world because of chemical bonds, multiquark hadron world should be
various due to the diversity of color flux-tube strucure. The well-defined the charged state $Z_c^+(3900)$ and dibaryon $d^*$ resonance
have been opening the gate of abundant multiquark hadronic world.

\acknowledgments
{C.R. Deng thanks Huanzhong Huang for hosting Deng's visit to UCLA. This research is partly supported by the NSFC under contracts Nos. 11305274,
11175088, 11035006, 11205091, and the Chongqing Natural Science Foundation under Project No. cstc2013jcyjA00014.}

\end{document}